\documentclass[%
	final
	]{issa}

\usepackage{graphicx}

\begin{document}

%%============================================================================
%% Title Page Information - Title, Authors, Affiliation, Abstract and Keywords
%%============================================================================
\title{SPAM over Internet Telephony and how to deal with it}

\author{Dr. Andreas U. Schmidt$^1$, Nicolai Kuntze$^1$, Rachid El Khayari$^2$}

\address{$^1$Fraunhofer-Insitute for Secure Information Technology SIT\\Rheinstrasse 75, Germany\\$^2$Technical University Darmstadt\\Germany}

\email{$^1$\{andreas.schmidt$|$nicolai.kuntze\}@sit.fraunhofer.de, $^2$rachid.el.khayari@googlemail.com}

\abstract{In our modern society telephony has developed to an omnipresent service. People are available at anytime and anywhere. Furthermore the Internet has emerged to an important communication medium.\\
These facts and the raising availability of broadband internet access has led to the fusion of these two services. Voice over IP or short VoIP is the keyword, that describes this combination.\\
The advantages of VoIP in comparison to classic telephony are location independence, simplification of transport networks, ability to establish multimedia communications and the low costs.\\
Nevertheless one can easily see, that combining two technologies, always brings up new challenges and problems that have to be solved.
It is undeniable that one of the most annoying facet of the Internet nowadays is email spam. According to different sources email spam is considered to be 80 to 90 percent of the email traffic produced.\\
Security experts suspect that this will spread out on VoIP too. The threat of so called voice spam or Spam over Internet Telephony (SPIT) is even more fatal than the threat that arose with email spam, for the annoyance and disturbance factor is much higher. As instance an email that hits the inbox at 4 p.m. is useless but will not disturb the user much. In contrast a ringing phone at 4 p.m. will lead to a much higher disturbance.\\
From the providers point of view both email spam and voice spam produce unwanted traffic and loss of trust of customers into the service.\\
In order to mitigate this threat different approaches from different parties have been developed. This paper focuses on state of the art anti voice spam solutions, analyses them and reveals their weak points.
In the end a SPIT producing benchmark tool will be introduced, that attacks the presented anti voice spam solutions.
With this tool it is possible for an administrator of a VoIP network to test how vulnerable his system is.}

\maketitle               

\section{Introduction}
In the following sections we will discuss the problematic of SPAM over Internet Telephony. The first section will deal with a general SPIT explanation and classification, followed by a scientific SPIT threat analysis.\\
In the second section the state of the art in SPIT prevention mechanisms will be presented 
and their weaknesses analysed. In the last section we will take a look at our SPIT 
benchmark tool (SXSM - SIP XML Scenario Maker) and how this tool can exploit the weaknesses 
of anti SPIT mechanisms.

\section{SPIT versus SPAM}
The focus of this paper is set on the topic of so called SPAM over Internet Telephony (SPIT). The first aspect to mention is, that although SPIT contains the phrase "SPAM" and has some parallels with email spam, it also has major differences. The similarity is that in both cases senders (or callers) use the Internet to target recipients (or callees) or a group of users, in order to place bulk unsolicited calls \cite{paper:SPITieee}. The main difference is that an email arrives at the server before it is accessed by the user. This means that structure and content of an email can be analysed at the server before it arrives at the recipient and so SPAM can be detected before it disturbs the recipient. As in VoIP scenarios delays of call establishment are not wished, session establishment messages are forwarded immediately to the recipients. Besides this fact the content of a VoIP call is exchanged not until the session is already established. In other words if the phone rings it is too late for SPIT prevention and the phone rings immediately after session initiation, while an email can be delayed and even, if it is not delayed, the recipient can decide if he wants to read the email immediately or not.\\
In addition to these aspects another main difference between spam and SPIT is the fact, that the single email itself contains information, that can be used for spam detection. The header fields contain information about sender, subject and content of the message. A single SPIT call in contradiction is technically indistinguishable from a call in general. A SPIT call is initiated and answered with the same set of SIP messages as any other call.

\section{Intuitive SPIT definition}
SPIT is described very similar in different publications and the descriptions can be summarised as  "unwanted" , "bulk" or "unsolicited" calls. In \cite{paper:SPITreach} e.g. SPIT is defined as "unsolicited advertising calls", which is already a special form of SPIT. In \cite{paper:SPITieee} SPIT is defined as "transmission of bulk unsolicited messages and calls" which is a more general definition than the first one, as it doesn't characterise the content and includes also messages. Note that with this definition it is not clear, if the term "messages" is used in order to generalise the type of messages that are sent (e.g. "SIP INVITE" or "SIP OPTIONS" messages) or, if it is used in order to include SPAM that is sent over Instant Messages (SPIM = SPAM over Instant Messages). The most precise definition is found in \cite{rfc:5039} where "Call SPAM" (as the authors call it) is defined as "a bulk unsolicited set of session initiation attempts (e.g., INVITE requests), attempting to establish a voice, video, instant messaging, or other type of communications session". The authors of \cite{rfc:5039} go even one step further and classify that "if the user should answer, the spammer proceeds to relay their message over the real-time media." and state that this "is the classic telemarketer spam, applied to SIP". We can easily see, that the presented definitions so far are very similar, but differ in their deepness.
\section{SPIT analysis}
The problem with the definitions above, is that they are either to specific or to general. In order to find a more precise definition, we have to analyse how SPIT is put into execution
and what the goal of the initiator of SPIT is.% and how does he or she achieve this goal?

In practice the initiator of SPIT has the goal to establish a communication session with as much victims as possible in order to transfer a message to any available endpoint. The attacker can fulfil this via three steps. 
First the systematic gathering of the contact addresses of victims. Second is the establishment of communication sessions with these victims and the third step is the sending of the message.

In the following we will not only discuss, why the process of information gathering is part of the SPIT process, but we will also see that it is the basis of any SPIT attack. In order to contact a victim, the attacker must know the SIP URI of the victim. We can differ permanent SIP URIs (e.g. sip:someone@example.com) and temporary SIP URIs (e.g. sip:someone@192.0.2.5).
\subsection{Information gathering}
At first we will take a look at information gathering of permanent SIP URIs. If an attacker wants to reach as many victims as possible he must catalogue valid assigned SIP URIs. The premisses for the \textbf{Scan attack} are the possession of at least one valid account and knowledge about the scheme of SIP URIs of the targeted platform (e.g. provider).\\
Let us assume the attacker has a valid SIP account at SIP provider "example.com" and he wants to scan the provider's network, in order to achieve a list of valid permanent SIP URIs. Let us also assume that the provider "example.com" distributes SIP URIs that correspond to the following scheme: The user name of the SIP URI is a phone number that begins with the digits "555" followed by 4 more random digits. All phone numbers from "5550000" to "5559999" are valid user names of this provider. As the attacker has now knowledge about all valid user names, he must find out which of them are already assigned to customers and which of them are not. The attacker can now step through the whole list of valid SIP URIs and send adequate SIP messages to each URI and receive information about the status of the tested URI. The simplest way is sending an INVITE message to each SIP URI and analyse the answer of the SIP Proxy. If the SIP URI is not assigned, the SIP Proxy may answer with a "404 Not found" response, if the SIP URI is assigned but the user is not registered at the moment, the SIP Proxy may answer with a "480 Temporarily unavailable" response and if the SIP URI is assigned and the user is registered, the call will be established and answered with a "200 OK" response. When the attacker has stepped through the whole list and marked all possible SIP URIs, he has a list of assigned SIP URIs, that can be used for future attacks. Note that it is not necessary that the scan attack must be fulfilled with an INVITE message, we just discussed this way as the simplest way, because it already leads to the desired session establishment. The attacker could also use an OPTIONS request or a REGISTER request and analyse the reaction of the Proxy. Mainly the implementation of the targeted Proxy decides on which message will grant the desired information. Some Proxies e.g. respond to all OPTIONS requests with a "200 OK" message, even in case of an invalid or unassigned SIP URI.\\
Now we will take a look at gathering of temporary SIP URIs. Temporary SIP URIs consist of the user name part and the host part. The user name part is usually a string or a phone number and the host part is the IP, where the endpoint can be reached directly. If an attacker has already generated a list of valid assigned SIP URIs, he now additionally needs the corresponding IP addresses of the SIP URIs. In some Proxy implementations the temporary SIP URIs are published in the "Contact" header of the response message to a request. In this case the desired information is achieved in the same way as the permanent SIP URIs. If the proxy does not provide the IP address in the SIP responses, the attacker must use a more complex method to achieve the desired information.
Let us assume this time that "example.com" is an Internet Service and VoIP provider. The provider assigns IP addresses of the range 192.0.2.5-192.0.2.155 to his customers and SIP URIs with the same scheme as described above (555XXXX). Let us assume The customers have hardphones (e.g. analogue telephone attached to VoIP ready router or Analog telephony adapter). With this knowledge the attacker can step through the list of IP addresses and try sending an adequate SIP request (INVITE,OPTIONS) directly to the endpoint (e.g. to UDP Port 5060) and analyse the responses in the same way as described above. The attacker can populate a list of temporary SIP URIs. Note that the temporary SIP URIs are only valid for a short time period (max. 24 hours), as customers are usually forced to disconnect their internet connection after a certain period. Although this procedure is harder to fulfil than the first one, it has the major advantage, that the attacker doesn't need valid accounts as premiss. Because the Proxy is not involved and SIP messages are sent directly to the victim, the attacker can use any SIP identity he wishes as source address. The client can not verify the identity, as nearly all existing implementations of clients accept SIP messages from any source.\\
Now that we have seen how lists of permanent or temporary SIP URIs can be achieved, we will discuss the usage of them.
\subsection{SPIT session establishment}
When the attacker has collected a large number of contact addresses, he can begin session establishment to the victims. Which list he must use (temporary or permanent URIs) depends on the communication infrastructure he wants to use. We can distinguish two possible ways of session establishment: The attacker can establish a session with sending an INVITE message via Proxy, which we can call \textbf{SPIT via Proxy} or he can establish a session with sending an INVITE message directly to the endpoint without involving the Proxy, which we can call {Direct IP Spitting}. For SPIT via proxy the attacker only needs a list of permanent SIP URIs and for Direct IP Spitting he needs the list of temporary SIP URIs. Again for SPIT via Proxy the attacker needs at least one valid user account and for Direct IP Spitting he doesn't need a valid account at all.
\subsection{SPIT media sending}
The last step of the SPIT process is the media sending after the session has been established. Which type of media is sent, depends on the scenario in which the SPIT attack takes place. The best scenario classification can be found in \cite{paper:SPITreach} and defines three types of SPIT scenarios:
\begin{itemize}
\item Call Centres: In Call centres a computer establishes a call to an entry of the catalogue and then dispatches the call to a call centre agent who will then talk to the callee.
\item Calling Bots: A calling bot steps through the list of gathered information, establishes a session and then sends a prerecorded message.
\item Ring tone SPIT: Some VoIP telephones come pre-configured in a way that they accept a special SIP header information called "Alert-info" which may contain an URL pointing to a prerecorded audio file somewhere on the Internet. Obviously, this can be used to play advertising messages before the call has even been accepted by the user just as the phone is ringing. An adaption of this method could be a SPIT attack where the attacker just wants to let the victims phone ring, in order to disturb the victim. In this special case no media is sent at all and the session is terminated as soon as the phone rings (e.g. when a "180 Ringing" is received). Obviously this is the most annoying facet of SPIT.
\end{itemize}
\subsection{SPIT summary}
As we can see now the SPIT process is very complex and has different aspects which have to be considered in order to develop countermeasures. The general definitions that we discussed in the first section are insufficient as a basis of discussion and do not cover all facets of the problem. In general we can say, that Spitting describes the systematic scanning of a VoIP network with the target of gathering information about available user accounts and the systematic session establishment attempts to as many users as possible in order to transfer any kind of message.

\section{SPIT countermeasures and their weaknesses}
In the following sections we will discuss state of the art SPIT prevention mechanisms in order to point out their advantages and disadvantages. The countermeasures are ordered by type and not by publication. As a matter of fact most publications define a set of countermeasures as a solution to mitigate SPIT. Nevertheless we will discuss every method on it's own and not the orchestration of different mechanisms. Note that only those techniques are listed, that have crystallised in research.
\subsection{Device Fingerprinting}
The technique of active and passive device fingerprinting is presented in \cite{paper:fingerprint} and is based on the following assumption:\\
Having knowledge about the type of User Agent that initiates a call, helps finding out whether a session initiation attempt can be classified as SPIT or not. The assumption is based on the analogy to e.g. HTTP based worms. As described in \cite{paper:fingerprint} these types of worms have different sets of HTTP headers and different response behaviour, when compared to typical Web browsers. So if we can compare the header layout and order or the response behaviour of a SIP User Agent with a typical User Agent, we can determine if the initiated session establishment is an attack or a normal call.\\
The authors describe two types of techniques that can be used for that purpose "Passive and Active Device Fingerprinting".
\subsubsection{Passive Fingerprinting}
The e.g. INVITE message of a session initiation is compared with the INVITE message of a set of "standard" SIP clients. If the order or appearance of the header fields does not match any of the standard clients, the call is classified as SPIT. The fingerprint in this case is the appearance and the order of the SIP header fields. The authors of \cite{paper:fingerprint} present a list of collected fingerprints of standard hard and soft phones.
\subsubsection{Active Fingerprinting}
User Agents are probed with special SIP messages and the responses are analysed and compared with the response behaviour of standard clients. The fingerprint in this case is the returned response code and the value of certain header fields. If the fingerprint doesn't match any of the standard clients, the call is classified as SPIT. The authors recommend the sending of specially crafted standard compliant and non compliant OPTIONS requests, in order to analyse the response behaviour of a client.
\subsubsection{Weakness of Device Fingerprinting}
The weakness of passive fingerprinting is described by the authors of \cite{paper:fingerprint} themselves. As passive fingerprinting only analyses the order and existence of the header fields of an INVITE message, an attacker simply needs to order the header fields in the same way as one standard client. In that case the passive fingerprinting mechanism can't detect the attack.\\
We can state nearly the same for active fingerprinting, as an attacker only needs to behave like one standard client when receiving unexpected or non standard compliant SIP messages. It is very simple for an attacker to develop an attacking SIP client that behaves exactly like a standard client, as he can use the same SIP Stack or imitate the behaviour of SIP Stack of a standard client. We can call this attack \textbf{Device Spoofing} and any attacker, who is able to spoof a device can nott be identified.\\
As Device Fingerprinting is discussed as a server side anti SPIT mechanism, it is useless against Direct IP Spitting as the clients don't have any chance to verify the fingerprint of the attacking client.\\
In the end we will take a look on practical issues of Device Fingerprinting. When we take a look at today's VoIP universe, we will find out that there exist a vast variety of hard- and softphones. Each of this phones has it's own SIP Stack and even within a product family header layouts and behaviour differ even between two versions of the same device. The result is, that an administrator who uses Device Fingerprinting in order to protect his system, must always keep the list of fingerprints up to date. Comparing the INVITE message of a caller with an old or incomplete fingerprint list, can lead to blocking the call although the call is not a SPIT call. Let us e.g. assume that a caller uses a standard client and that the manufacturer sends out a firmware upgrade, that makes major changes to the SIP Stack. Any calls of this user are blocked or marked as SPIT, until the administrator of the VoIP network updates the fingerprint list and this procedure will repeat any time a new firmware version is rolled out or new clients are released.\\
Taking it even one step further, we can see, that as more and more clients and versions are released, the fingerprint list will become wider and wider and in the end nearly any combination of e.g. header fields will be present in the list. The main problem of device fingerprinting is that it is derived from a HTTP security technique. In that scenario only few clients (web browsers) from few developers exist, in contradiction to the VoIP world.
\subsection{White Lists, Black Lists, Grey Lists}
The White List technique is presented e.g. in \cite{paper:SPITreach} \cite{rfc:5039} and works as follows: Each user has a list of users that he accepts calls from and any caller who is not present in the list will be blocked. In addition the private White Lists can be distributed to other users. If e.g. a caller is not present in the White List of the callee, White Lists of other trusted users can be consulted and their trusted users (up to a certain level), however this technique needs additional mechanisms.
Black Lists are the contradiction of White Lists and contain only identities, that are already known as spammers. Any call from a caller whose identity is present in the callee's Black List is blocked. Even Black Lists can be implemented as distributed Black Lists, where a callee can consult the Black Lists of other users.
Grey listing works as follows: On initial request of an unknown user (not in White List) the call is rejected and the identity is put on the Grey List. As stated in \cite{paper:SPITreach} in case the caller tries calling back within a short time period, the call will be accepted. An adaption of this technique is described in \cite{rfc:5039} as Consent Based Communication. In case of Consent Based Communication the call of an unknown caller is initially blocked and put on the Grey List. The callee can consult the Grey List and decide, if he will accept future calls from this identity or block it permanently.
\subsubsection{Weaknesses of White Lists, Black Lists, Grey Lists}
Black Lists can not really be viewed as a SPIT countermeasure, because additional methods are needed to classify a caller a Spitter. A Black List on server side would require e.g. statistical methods for classifying a caller as Spitter. In case of a client side Black List, the user must mark a caller as a Spitter, e.g. after receiving an initial SPIT call from this caller.
Both server side and client side Black List are very useless against Direct IP Spitting for different reasons. Server sided Black Lists are bypassed by Direct IP Spitting, because the SIP messages are sent directly to the client. Client sided Black Lists are circumvented by Direct IP Spitting, because the caller can take on any identity in order to place calls. So if one identity is blocked he can simply switch the Identity. We can call this attack \textbf{SIP Identity Spoofing} and any attacker who can spoof SIP identities, can easily bypass Black Lists.\\
White Lists are at first sight harder to circumvent than Black Lists, because the attacker has no knowledge about the entries of the White List of the victim. So even if he wants to spoof an identity, the attacker doesn't know which identity he must take on, in order to place a successful call. In case of Direct IP Spitting the attacker could simply try out all existing accounts with a brute force attack until he finds out which identities are not blocked. A less exhausting procedure can be performed in case of distributed or imported white lists \cite{paper:SPITreach}. In that scenario the attacker needs one valid account. After adding the victim to the attacker's white list, he can now select that he wants to import the white list of the victim. So he can get access to all entries of the victim's white list and can spoof these identities e.g. in a Direct IP Spitting attack.
The Grey List mechanism can be bypassed the same way as White List mechanisms, as it just represents a mechanism that allows first time contact. All in all we can say, that any attacker who is able to perform SIP Identity Spoofing, can bypass Black Lists, White Lists and Grey Lists.\\
In the end we will take again a look at the practical side of the presented mechanisms. The concepts of Black, White and Grey Listing are derived from the Instant Messaging world, where it is a matter of course, that users first ask for permission, before they are added to another user's buddy list and only buddies can communicate with each other. When a user receives a communication request, he receives the profile of the other user containing e.g. nick name, email address, full name or even profile photo. On basis of this information, the user can decide and is able to decide, if he wants to accept messages in future from that party or not. Taken to the VoIP scenario this mechanism seems very impractical as the introduction problem has to be solved. Let us assume e.g. an employee of a bank wants to call one of his customers. In case of white listing the call can not be successfully routed to its target, as customers usually don't have the phone numbers of employees of their home bank listed in the White List. The decision basis for accepting or rejecting a call is simply the phone number that is sent by the caller. If the call is rejected at first (Grey listing) the callee must decide if he wants to accept future calls and he must base this decision on the phone number. We can easily see that this fact is very impractical.
\subsection{Reputation Systems}
Reputation based mechanisms are described in \cite{rfcdraft:feedback} or in \cite{rfc:5039} and can be summarised as follows:
After receiving a call, the callee can set a reputation value for the caller, that marks this caller as Spitter or not. This reputation value must be assigned to the identity of the caller and can be used for future session establishment requests. This technique can be used e.g. as attachment to Grey listing \cite{rfc:5039} in order to provide a better decision basis. The authors of \cite{rfcdraft:feedback} explain that the user feedback can be used additionally for calls that were not detected by other SPIT preventing components. The way the reputation value is generated can differ. The SPIT value can be e.g. an additional SIP header, or included in a special error response code or distributed via SIP event notification mechanism. Reputation systems can be either based on negative or positive reputation values. This means that in first case only Spitters are marked with negative values or in second case "normal" callers are marked with positive values.\\
An adaption of this method can be found in \cite{paper:p2pavs} where user feedback is combined with statistical values in order to calculate a reputation value. The reputation value is e.g. composed of a value representing the number of times an identity occurs in other users' Black Lists, call density, call length or similar statistic values. The assumption behind this approach is that the calculated value will differ much between "normal" users and spitters.
\subsubsection{Weakness of Reputation Systems}
Reputation systems that are based on negative reputation can be bypassed in same way as Black Lists \cite{rfc:5039}. A user with a negative reputation can be viewed as globally blacklisted as his calls are blocked e.g. for any user (this depends on the policy that is used). Nevertheless an attacker that is black listed simply needs to gain access to a new "clean" account. In case of a SPIT value as SIP header, the SPIT value can be spoofed by the attacker (e.g. with Direct IP Spitting) and we can call this attack \textbf{SIP Header Spoofing}.
The attacker can simply set or change values of header fields, when he uses Direct IP Spitting.\\
In addition an attacker can create several accounts with the aim of pushing the SPIT value of one account up or down (depending on implementation). This attack can be called \textbf{Reputation Pushing or Pulling}.\\
Again we will also take a closer look of practical issues of the anti SPIT mechanism. At first we must admit, that Reputation systems are more auxiliary features than SPIT blocking mechanisms. The reason for this argumentation is, that the user must classify a call as SPIT via a button or by entering a value. This value is used for future decisions on that SIP identity. So initially SPIT is not prevented by this technique. Then the SPIT value of an identity has to be shown to callees, so that they can decide about accepting or rejecting the call. Let us assume a Spitter has achieved a SPIT value or SPIT probability of e.g. thirty percent and then calls a victim. What should happen now? When the call is forwarded to the user and the value is e.g. shown in the display of the callee's phone, he can decide to accept or reject the call on a better decision basis. The problem is that anyhow his phone rings and that is what should be prevented. He could have just picked up the call and listened the first 5 seconds to know that it is SPIT. So the SPIT value didn't just add one percent of benefit. On top of this fact attackers could misuse the scoring system and create enough accounts in order to threaten "normal" users with collectively giving them negative reputation \cite{rfc:5039}
\subsection{Turing tests, Computational Puzzles}
Turing test are tests where the caller is given a challenge, that a human can solve easily and that is hard to solve for a machine. Therefore Turing tests or CAPTCHA (Completely Automated Public Turing test to tell Computers and Humans Apart) are tests, that countermeasure Calling Bot attacks in VoIP scenarios. Turing tests in VoIP scenario work as follows: On initial call establishment attempt, the caller is transferred to an interactive System where he is challenged with a task e.g. dialing 5 digits that he is hearing (so called Audio CAPTCHA). While the numbers are read out background music or any other kind of noise is played, so that speech recognition systems can't be used to solve the task. A human caller in contradiction will solve the task without difficulties and only if the task is solved, the call will be forwarded to its destination. Turing tests can be used in combination with white lists, solving the introduction problem as described in \cite{rfcdraft:audiocaptcha}.\\
Computational Puzzles seem at first sight very similar to the Turing tests concept. As described in \cite{rfcdraft:comppuzz} a SIP Proxy or User Agent Server can request from a User Agent Client (caller) to compute the solution to a puzzle. The goal of this method is to raise CPU costs of a call and so reduce the number of undesirable messages that can be sent. Turing test in contradiction have the goal to block non-human callers, as described above. According to \cite{rfcdraft:comppuzz} the puzzle, that has to be solved, could be finding a pre-image that will SHA1 hash to the correct image. This means that the UAC will be challenged with a SHA1 hash of a value and the UAC must find out (by computing it) which value has been hashed.
\subsubsection{Weakness of Turing tests and Computational Puzzles}
Turing tests seem at first sight very effective for SPIT prevention in combination with white lists, but nonetheless have weak points. The first approach of bypassing Audio CAPTCHA is relaying the CAPTCHA to human solvers. An attacker could pay cheap workers, who are only hired to solve Audio CAPTCHA. In countries with cheap labour this would raise the costs per call only marginally \cite{rfc:5039}. In order to reduce the costs, an attacker could even e.g. set up an adult hotline and could dispatch Audio CAPTCHA to the customers of this service. This technique is known from visual CAPTCHA where the images from CAPTCHA protected sites are copied and relayed to a high traffic site owned by the attacker. All in all we can state, that an attacker who can detect CAPTCHA and relay it to human solvers is able to bypass Turing tests and we can call this attack \textbf{CAPTCHA Relay Attack}.\\
Computational Puzzles can not be viewed as SPIT prevention mechanisms, as attackers usually possess high computational power. So Circumventing a system protected by Computational Puzzles, doesn't even demand a special attack. The attacker just needs sufficient CPU power.\\
In the end again we will take a look at some practical issues of the described techniques. As far as Turing tests are concerned, we can see, that this method is very intrusive. User Interaction is forced every time a caller is not present in the White List of a callee.\\
The difficulty with Computational Puzzles is, that different VoIP endpoints have different abilities in computational power. So if the task is to hard to solve (consumes too much CPU power), session establishment will be delayed very much for e.g. a low-end cell phone, while attackers with high CPU power PCs won't be concerned much. With this fact Computational Puzzles are very ineffective and contra productive, as they only bother "normal" users.
\subsection{Payments at risk}
Payments at risk mechanisms can be used in order to demand payment from an unknown caller. In \cite{rfc:5039} this technique is described as follows: If user A wants to call user B, he must first send a small amount of money to user B. When User B accepts the call and confirms that the call is not a SPIT call, the amount will be charged back to user A. With this technique it is possible to raise costs for SPIT callers while keeping "normal" calls cheap. In \cite{rfc:5039} it is described as an auxiliary technique that solves the introduction problem of White lists, tis means, that payment is only required for callers who are not on the White list of callee. In general the payment could be demanded for every call, but this would make the telephony service more expensive.\\An adaption of this method is described in \cite{dipl:unipotsdam}, here the Payment technique is used in combination with a SPIT prediction value that is computed at server side. If the SPIT likelihood is high the call is rejected, if the SPIT likelihood is small the call is forwarded to the callee and if the SPIT likelihood value is in between payment is demanded automatically. Only if the payment is fulfilled the call will be forwarded to its target. The difference between the two approaches is, that in the first case the payed amount is only charged back for non SPIT calls and in the second case, callers who reject payment are treated as Spitters.
\subsubsection{Weakness of Payment at risk}
In which way Payment at risk can be bypassed depends mainly on the way it is implemented. As described demanding payment for each call won't be very realistic, because this would require a high administrative overhead and more costs for service providers. Let us assume Payment at risk combined with White listing as in the first example, so that payment is only required for callers that are not present in the callee's White List. In this case a caller could simply spoof identity as described in the section about White List.\\ In the second scenario, where Payment at risk is combined with a Reputation system, the attacker just needs to achieve an adequate reputation value, as described in the corresponding section.\\ Let us even assume, that Payment at Risk is used for every call. Even In that case an attacker could circumvent it, by impersonating as another user, so that he can establish calls and shift the costs on to "normal" customers. In which way this kind of \textbf{SIP Identity Hijacking} attack is fulfilled is an other question and out of scope for now.\\ Besides the technical aspects, practical issues of Payment at Risk are numerous. At first the relative high costs, that are required for micropayment will must be viewed, the inequities in the value of currency between sender and recipient \cite{rfc:5039} and the additional interactions that a user must take (e.g. confirming a call from an unknown party as non SPIT).
\subsection{Intrusion Detection Mechanisms, Honey phones}
Intrusion Detection Systems are (generally described) systems, that can be used for detection of any kind of abnormal behaviour within a e.g. network and so reveal attacks. An implementation of this technique is presented in \cite{paper:intrusion} based on the Bayes inference approach combined with network monitoring of VoIP specific traffic. The Intrusion Detection System is designed as a defence mechanism against different VoIP specific attacks including scan attacks and SPIT attacks. For every attack a conditional probability table (CPT) is defined for variables such as request intensity, error response intensity, parsing error intensity, number of different destinations, max number of dialogues in waiting state, number of opened RTP ports, request distribution and response distribution. Let us look at e.g. the CPT for the number of different destinations variable: For a SPIT attack the likelihood of having more than 7 different destinations is set to 1 and the likelihood of having up to 7 different destinations is set to 0. The concept behind this technique is, that the different attacks affect these variables in different ways, e.g. a SPIT attack usually has a higher probability of a higher number of destinations than normal traffic. So a belief of a network trace can be calculated with the aid of likelihood vectors that were defined in the CPT. In the end the trace can be categorised as an attack or normal trace (refer to \cite{paper:intrusion} for detailed description).\\
Honey phones can be used as part of an Intrusion Detection Systems as described in \cite{paper:SPITreach} \cite{paper:heoneyhol} and can be viewed as VoIP specific Honeypots. A Honeypot represents a part of a network that is not accessible by "normal" users and therefore any access to the honeypot can be viewed as an attack. VoIP specific honeypots can be used in order to detect Scan attacks or SPIT attacks. As described in \cite{paper:heoneyhol} the Honeypot is implemented as a complete parallel VoIP infrastructure, that is logically and physically separated from the normal network and so simulates a whole VoIP network. Let us assume a Scan attack as described earlier. When the attacker sends e.g. OPTIONS or INVITE requests to valid assigned permanent URIs they are forwarded through the normal SIP network (Proxy, UAC), but when the attacker tries to send an OPTIONS request to an unassigned or invalid SIP URI the request will be forwarded to the Honeypot, where the requests can be monitored and treated adequately. The authors of \cite{paper:heoneyhol} propose call analysis in order to determine attack characteristics, interaction with the originator in order to determine the source of the attack and blocking of the calls, as adequate treatment. The monitoring system of this approach works as follows: A day is divided into sections of specified time (e.g. one hour). For each section a predefined metric is calculated (e.g. number of calls, number of different recipients, average duration of a call) matching predefined events (e.g.call). In the learning phase (e.g. a month), daily statistics are
built to extract a long term account profile (e.g. daily average of the number of calls for each section). In the detecting stage (e.g. a day), a short term profile is compared to the long term
one by using an appropriate distance function (e.g. Euclidean distance, quadratic distance, Mahalanobis distance). A recent profile which is quite different from the long term one indicates
possible misuse. Another method is to study non stationary features of an account, for example the distribution of calls over all callees or the shape of the callees' list size over all dialed calls. By comparing changes of a distribution over the time by using of an appropriate distance function (e.g. Hellinger distance), sudden bursts may be detected and treated as abnormalities \cite{paper:heoneyhol}.
\subsubsection{Weakness of Intrusion Detection Mechanisms, Honey phones}
Intrusion Detection Systems base on the assumption, that the characteristics of attacks differ much from characteristics of normal calls. At first sight this assumption seems logic, as e.g. within a SPIT attack, the attacker calls hundreds or thousands of victims within an hour, while a normal user wouldn't even send out one percent of this amount of calls. Nevertheless the attacker has two possibilities in order to bypass detection by an Intrusion Detection System. The first is to align his behaviour with the behaviour of normal users, e.g. adjust the call rate to 5 calls per hour. Obviously this technique is hard to fulfil, because this would make an attack very inefficient as it would consume too much time, but on the other hand the goal of a spitter is not to reach as much users as possible within the shortest time period. Reaching e.g. thousand users with a call rate of 5 calls per hour would take approximately 8 days. We can call this technique \textbf{Call Rate Adaption}, this means that an attacker is able to adjust his call rate (e.g. number of calls per time slot, number of simultaneous calls). As the call rate is not the only variable that is used in order to detect abnormal behaviour an attacker can use a second technique in order to not be detected by Intrusion Detection Systems. The attacker can use different accounts for his attacks, so that statistic values are spread over several accounts. Let us assume that an attacker has one hundred valid user accounts. With this amount of accounts he can partition the targeted user accounts into one hundred groups and use only one account per group. The users from group one are only called with account one and so on. It is harder for a monitoring system to detect attacks that are originated from different sources, as there must be a technique to correlate partial attacks to one complete attack. This technique can be called \textbf{Account Switching}, as the attacker switches the used account while he is performing an attack.\\
Honeypots are very effective against scan attacks as anyone who tries to reach invalid or unassigned identities, will be trapped and so Honeypots are very effective against SPIT. When the Spitter can't scan the network for assigned and unassigned numbers, he is forced to view all numbers as assigned. When he views all numbers as assigned, he will sooner or later step into the trap, because he will establish calls to endpoints, that are part of the Honeypot. Nevertheless attackers can trick the Honeypot mechanism with \textbf{SIP Identity Hijacking}. When an attacker impersonates the accounts of normal users and then performs SPIT attacks with this normal accounts, he will access end points in the Honeypot system with normal accounts. So the assumption that accesses to the Honeypot are only established by attackers is lapsed.\\
In the end we will take again a look at the practical issues of the presented solutions. The practical problem with intrusion detection systems in general is, that they base on statistical assumptions that are not verified. The questions that has to be solved is: Where is the borderline between normal usage and abnormal usage? The publishers state that statistical values are assumed or derived from attack characteristics, but in order to reduce the rate of false negative and false positive classifications, the knowledge basis must be precise. So we can say that what we lack, is knowledge of SPIT characteristics as we nowadays can't really distinguish SPIT from normal traffic unless the SPIT attacks are excessive.
Honeypots have the disadvantage, that they only detect access to invalid or unassigned accounts, this means that an attacker who only accesses valid accounts won't be handled by a honeypot. 
\subsection{Summary}
We can finally say, that we have seen SPIT countermeasures with different weak points. All of the presented ideas have technical and practical weak points, that can be exploited by attackers in order to circumvent these technique. An attacker who is able to perform 
Device Spoofing, SIP Identity Spoofing, SIP Header Spoofing, Reputation Pushing or Pulling,
CAPTCHA Relay Attack, SIP Identity Hijacking, Call Rate Adaption, and Account Switching
has a good repertoire, that enables him to bypass any of the presented techniques or combinations of them. An attacker now needs a tool that aggregates the presented attacks.
\section{SIP XML Scenario Maker}
In the following sections we will introduce our SPIT producing benchmark tool, that implements the presented attacking techniques.
\subsection{Technical Basis}
SXSM is based on SIPp developed by HP \cite{homepage:sipp}. SIPp is an Open Source test tool and traffic generator for SIP. SXSM expands SIPp with the ability to quickly create custom SIP scenarios via a graphical user interface (GUI), execute created scenarios and evaluate the result of the execution. The functionality  is fulfilled by two different editing modes and one execution mode.
\subsection{Message Editor}
The message editor delivers the basis functionality, the ability to create custom SIP Messages. SIP messages are the smallest elements of SIP scenarios, as SIP scenarios are sequences of SIP messages.\\ Within the message editor, the user can configure the layout of each and every SIP message, that can be used in the scenario editor (explained later). Additionally the messages can be grouped into sets, so that the user can create e.g. two differently composed INVITE messages and put one into set A and one into set B. Later the user can distinguish the two INVITE messages, because they are in different sets. In the message editing mode the user can even compose or modify existing SIPp control messages (e.g. pause commands) that can be used in order to control the behaviour during execution. SXSM comes preconfigured  with a set of standard messages that can be used as orientation in order to compose own messages.
\subsection{Scenario Editor}
The scenario editor is the core element of SXSM. In this mode the user can create SIP scenarios, based on the bricks created in the message editor. Even the scenarios can be grouped in different sets. The user must chose a name for a scenario and can then select from the list of messages, the messages he wants to add to the scenario and in which order they should appear. Afterwards the user can edit the scenario, that is presented as an XML file, in detail. Let us assume the user wants to create a scenario where an INVITE message is sent, then a "100 Trying" is received, then a "180 Ringing" is received then a "200 OK" is received. In this case the user simply selects these messages and adds them to the scenario, saves the scenario file and work is done.
\subsection{Shoot Mode}
The shoot mode represents the execution mode. In this mode the user can put previously created SIP scenarios into a sequence, execute them one after the other and evaluate the results presented.\\The user selects scenarios from the scenario list and adds them to the shoot list, then he configures the call rate (calls per time period), then he enters information about the target (remote IP, remote port) and about himself (local IP, local port). After this information has been provided, he must provide information about the SIP identities, that should be used for the execution. The user can choose, if he wants to use fixed values for both source and target SIP URI or inject values from an external CSV file. In first case he must provide a user name for the targeted SIP URI and this user name will be used for each and every call, that is executed through the shoot list. If the user wants to inject values from an external CSV, he must specify the location of the file. After all parameters have been set, the user can execute the shoot list. SXSM then feeds SIPp with the input data and waits until the scenarios are executed. When the execution is fulfilled, SXSM evaluates the exit codes, that were generated by SIPp. The following exit codes are considered:
\begin{itemize}
\item 0: All calls were successful
\item 1: At least one call failed
\item 97: exit on internal command. Calls may have been processed. Also exit on global timeout
\item 99: Normal exit without calls processed
\item -1: Fatal error
\end{itemize}
Based on the exit code a success rate is calculated and displayed. If e.g. all scenarios of the shoot list completed with exit code "0" then the success rate is "100". Additionally log files will be presented for each scenario, that can be used for debugging purpose in case of failed scenarios.
\subsection{Using SXSM as attack tool} 
As SXSM is implemented within a very broad and modular context, it can be used for all SIP testing purposes and in special as a SPIT producing attack tool. In the following we will discuss how the different attacks, that were presented in the previous sections, can be put into practice.
\subsubsection{Device Spoofing}
The Device Spoofing attack is an attack, that has two facets. As we discussed earlier device fingerprints can be derived from the layout of the SIP messages or from the behaviour. The layout of SIP messages can be manipulated within the message editor of SXSM. If a user wants to imitate the message layout (presence and order of SIP headers) of a device, he simply needs to create a new set of SIP messages, name the set after the device and create the desired SIP messages, according to the wished layout.\\
The behaviour of the client can be manipulated within the scenario editor. The ability to create scenarios that contain branch points eases this process. A scenario can then contain a section for every message that can be received. The user must only put tests into the scenario with the scheme "if message x is received jump to section y".
\subsubsection{SIP Identity Spoofing}
Identity Spoofing in its simple form is provided by inserting the wished SIP URI in the "From" and "Contact" header of the SIP messages. If the user wants to inject the SIP URI from an external CSV file he must specify this in the scenario file. The user simply needs to put the expression "[fieldn]" where n represents a number (the column of the CSV file) at the position where the user name is usually placed in the "From" or "Contact" header according to the following scheme:\\
\textit{From: [field0] $<$sip:[field1]@[local\_ip]:[local\_port]$>$;}\\
Here the first column of the "caller.csv" file contains the name of the identity and the second column contains the user name part of the SIP URI.
\subsubsection{SIP Header Spoofing}
SIP Header Spoofing can be fulfilled with two different approaches. The first one is to create custom SIP messages with the message editor and set headers and header values as wished. The second is using standard SIP messages and change values of headers with the detail view of the scenario editor. The first one should be preferred, if the user wants to create a lot of scenarios with the same header value and the second variant should be used, if the user wants to tweak values only once in a while.
\subsubsection{Call Rate Adaption}
The Call Rate Adaption attack can be fulfilled within the shoot mode. The user just needs to add a scenario to the shoot list and adjust the values for call rate. He can put e.g. "Scenario X" into the shoot list and set the call rate to 10 times per second and stop as soon as 100 calls have been finished. With this method it is possible to control in detail how many times in what time period a scenario is executed. As one and the same scenario can be present several times in the shoot list, the user can define the behaviour very precisely. So he can e.g. determine that "Scenario X" should be executed 100 times with a call rate of 10 calls per second and then 20 times with a call rate of 2 calls per second. Note that the phrase "call" means one pass of scenario from beginning to end and does not mean that a call is actually placed. A scenario could e.g. consist of sending an OPTIONS request and receiving the answer.
\subsubsection{Account Switching}
The Account Switching attack is a special form of SIP Identity Spoofing and can be fulfilled by providing an external CSV file with appropriate data. Let us assume the user wants to place 100 calls to hundred different targets with ten different SIP identities. The CSV file for the callee should contain hundred rows and each row should contain the user name of the targeted URI. The CSV file for the caller should contain 10 rows and each row should contain one of the ten user names, that should be used as source. Including the SIP identities for the caller can be fulfilled with the mechanism described in the section about SIP Identity Spoofing. Including the SIP identities of the target can be configured in the shoot mode by selecting the external CSV file as target.
\subsubsection{Reputation Pushing or Pulling}
Reputation Pushing or Pulling is very dependent of the implementation, but can be fulfilled in a generic way. The user simply needs to create two scenarios. One, that sends out a call and one, that receives a call. The receiving scenario should include e.g. a positive reputation value into the BYE message. Note, that this is the point where it is implementation specific, as it depends on the implementation of the Reputation System where the reputation value must be put. Then the user must launch two instances of SXSM and shoot out the calling scenario with one instance and the receiving scenario with the other instance. Combining this technique with Account Switching for the call receiving side can lead to the desired effect of Reputation Pushing or Pulling.
\subsubsection{SIP Identity Hijacking}
SIP Identity Hijacking is again very implementation dependant. but we can take a look at a simple attack derived from \cite{homepage:attack}. The registration hijacking attack is presented as follows:
\begin{enumerate}
\item Disable the legitimate user's registration. This can be done by:
\begin{itemize}
\item performing a DoS attack against the user's device
\item deregistering the user (another attack which is not covered here)
\item Generating a registration race-condition in which the attacker sends repeatedly REGISTER requests in a shorter timeframe (such as every 15 seconds) in order to override the legitimate user's registration request. 
\end{itemize}
\item Send a REGISTER request with the attacker's IP address instead of the legitimate user's
\end{enumerate}
With SXSM this process could be put into practice, by creating scenarios for each of the presented steps and execute them one after the other, but as the Session Hijacking attack has a variety of aspects, that have to be considered we will not discuss this matter in detail for now.
\subsubsection{CAPTCHA Relay Attack}
The CAPTCHA Relay Attack can be fulfilled with the Third Party Call Control (3PCC) mechanism. With this mechanism it is possible for SIPp (and therefore for SXSM) to create a communication session with several remote endpoints and so relay e.g. calls. The procedure is fulfilled as follows. The Attacker calls the victim, who sends the Audio CAPTCHA. In response the
attacker calls human solver and "REFER"s the victim to him. As the result the victim 
accepts and the human solver solves CAPTCHAs.
As this attack is as good as the used CAPTCHA detecting algorithm, the technique must be adapted to future implementations of both CAPTCHA generators and detecters.
\section{Conclusion}
This paper described the main aspects, that should be considered by developers of anti SPIT solutions. We saw, that a precise problem definition is mandatory for SPIT research, as we can only cover all aspects, if we clearly know which problems have to be faced. Then we saw, that state of the art anti SPIT mechanism still embody both technical and practical weak points, that can be viewed as vulnerabilities. In the last part of this paper we got an impression of how this vulnerabilities can be abused by attackers, with a simple attack tool, whose power lays in full control over the SIP protocol. The next step of research should consider the presented attacks and develop mechanism that blank out these vulnerabilities.

\end{document}